\documentclass[%
 aip,
 jmp,%
 amsmath,amssymb,
 reprint,%
]{revtex4-1}

\usepackage{graphicx}
\usepackage{dcolumn}
\usepackage{bm}

\begin{document}

\title{Determination of the critical growth rate and growth temperature for group-III elements segregation using two exchanges Kinetic Model}

\author{M. Arjomandi}
\affiliation{%
Department of Electronic Engineering, Islamic Azad University Arak Branch, 38361-1-9131 Arak, Iran
\email{m.arjomandi92@iau-arak.ac.ir}} %

\author{P. Hosseinzadeh}
\affiliation{%
Electronic Research Center, Iran University of Science and Technology, 1684613114 Tehran, Iran
}%
\author{S. Dadgostar}
\affiliation{%
Electronic Research Center, Iran University of Science and Technology, 1684613114 Tehran, Iran
}%

\author{V.R. Yazdanpanah}
\affiliation{%
Electronic Research Center, Iran University of Science and Technology, 1684613114 Tehran, Iran
}%

\author{S. Mirzakuchaki}
\affiliation{%
Electronic Research Center, Iran University of Science and Technology, 1684613114 Tehran, Iran
}%

\date{\today}

\begin{abstract}

Segregation of group-III elements during the molecular-beam epitaxy growth of III-V compounds leads to a non-abrupt interface. The composition asymmetry in the structures such as quantum wells, quantum dots, and superlattices, in turn, leads to the non-abrupt electronic band alignments that changes the optoelectronic properties of those quantum structures. We have studied the concentration profile of the group-III atoms for different growth parameters using two exchanges Kinetic Model and have determined the critical growth temperature and growth rate regions for the growth of structures with less than 10$ \% $ segregation of group-III atoms.  

\end{abstract}

\maketitle

High-performance optoelectronic devices such as lasers, light emitter diodes (LED), solar cells, and detectors can be fabricated based on the complex quantum structures of III-V semiconductors. Quantum structures such as quantum wells (QW), quantum dots (QD), and superlattices can be grown with high quality using molecular-beam epitaxy (MBE).
However, the interface composition abruptness is a challenge due to the surface atom intermixing. \cite{Dev, Moison}. 

For the growth temperatures below 600~$ ^{\circ}$C, atomic arrangement in the crystal is determined by surface or near-surface processes and atoms cannot rearrange after burial. However, due to the surface mobility, atoms can displace on the growing surface. The increase of the surface mobility leads to the elimination of the growth defects but, it may cause the so-called "surface segregation", which is the exchange between the sub-layer atoms with the impinging atoms on the growing surface. Surface segregation of atoms is driven by the differences in their binding and elastic energies \cite{Dehaes}. Several experimental and theoretical studies \cite{Moison,Dehaes,Muraki,Renard,Sozykin,Gerardi} indicate that the group-III atoms with weaker bond strength segregate to the surface, and therefore, it is expected to see more segregation for In atoms in comparison with Ga atoms for similar growth conditions\cite{Moison}. Experimental results on the well-known InGaAs/GaAs system show an Indium surface enrichment \cite{Martini, Riposan}. A similar behavior has been observed for In atoms surface segregation in InGaAs QDs embedded in GaP matrix \cite{Robert}. In AlGaAs/GaAs QWs, a composition asymmetry has been reported at the normal interface (AlGaAs on GaAs) that is associated with the Ga atoms segregation \cite{Braun,Etienne,Jusserand}. The results also indicate the intermixing of Al and Ga atoms at the GaSb/AlSb normal interface \cite{Gerardi,Bocchi,Lomascolo,Massies}.

Group-III atoms segregation increases with increasing the surface mobility and the growth conditions. The increase in the growth temperature increases the surface mobility and thus a stronger surface segregation of atoms is expected \cite{Muraki}. The surface mobility, on the other hand, decreases with increasing the growth rate and this leads to a lower segregation \cite{Dehaes}.
Segregation of group-III atoms causes a composition asymmetry at the interface of the heterostructures. This, in turn, results in the change of the electronic band alignment of the quantum structures from a symmetric to an asymmetric well, which alters the optoelectronic properties. Therefore, to design an optoelectronic device based on the III-V semiconductor structures, it is important to predict the concentration profile of the segregating atom for the different growth parameters such as growth temperature and growth rate. Kinetic Monte Carlo (KMC)\cite{Khazanova1} and two exchanges Kinetic Model (we call it Kinetic Model) for segregation\cite{Dehaes,Magri,Magri02} are two theoretical models that are widely used to calculate the profile of the segregating atoms. 
Using KMC, the whole growth processes can be simulated based on the short range surface diffusion of adatoms which exponentially depends on the activation energy for the surface diffusion\cite{Khazanova1}. Consequently, the concentration of constituent atoms can be calculated for each monolayer(1~ML = half of the lattice constant)\cite{Khazanova1}. 
Kinetic Model, on the other hand, is a kinetic thermodynamic model, where the profile of atoms concentrations is calculated based on the probability of the exchange of atoms on the surface with the atoms in the underlying layer\cite{Dehaes}. 

Several studies have been conducted based on the Kinetic Model to predict the effects of the growth conditions on the surface segregation of both group-III and group-V atoms\cite{Dehaes,Magri,Magri02}. However, the growth window for growth temperature and growth rate has not been determined. In this work, we have studied the segregation rate of group-III elements with respect to different growth temperatures and growth rates and have determined the growth window that can be interesting from the practical point of view.

The Kinetic Model simulates a layer by layer growth mode of an $A_{x}B_{1-x}C$ III-V alloy on a $BC$ substrate. Here, $A$ and $B$ are the group-III elements and $C$ belongs to the group-V. The exchange is considered between the atoms on the uppermost layer (surface) and in one layer below the surface (bulk). The exchange process occurs when atom-A overcomes a barrier energy of $E^{b\rightarrow s}_{A/B}$ to move from bulk to the surface. The inverse exchange also happens when atom-A on the surface overcomes the barrier energy of $E^{s\rightarrow b}_{A/B}$ and moves into the bulk. The exchange rate, therefore, is given by\cite{Dehaes,Magri}:
\begin{equation}
P^{b\rightarrow s}_{A/B}=\nu exp(\dfrac{-E^{b\rightarrow s}_{A/B}}{k_{B}T})
\end{equation}

and the inverse exchange rate is: 
\begin{equation}
P^{s\rightarrow b}_{A/B}=\nu exp(\dfrac{-E^{s\rightarrow b}_{A/B}}{k_{B}T})
\end{equation}
Where, $\nu = 10^{13} Hz$ is the atomic vibration frequency, $T$  is the growth temperature, and $ k_{B} $ is the Boltzmann constant.
Therefore, segregation driving force ($E_{s} $) is determined as:
\begin{equation}
E_{s} = E^{s\rightarrow b}_{A/B} - E^{b\rightarrow s}_{A/B} 
\end{equation}

Assuming that the segregation is only due to the exchange processes, the balance of the incoming and leaving atoms to the surface gives the evaluation of the number of atom-$A$ on the surface\cite{Dehaes}:
\begin{equation}
\dfrac{dX^{s}_{A}(t)}{dt} = \Phi_{A} + P^{b\rightarrow s}_{A/B} X^{b}_{A}(t) X^{s}_{B}(t)-P^{s\rightarrow b}_{A/B} X^{s}_{A}(t) X^{b}_{B}(t)
\end{equation} 

Here, $ \Phi_{A} $ is the impinging flux of atom-$A$, $ X^{s}_{A (or B)}(t) $ is the concentration of atom-A(or B) on the surface at any time ($t$), and $ X^{b}_{A (or B)}(t) $ is the concentration of atom-A(or B) at any time ($t$) in the bulk. Variation of atom-$A$ concentration in the time interval of $dt$ ($ \dfrac{dX^{s}_{A}(t)}{dt} $), therefore, is equal to the summation of $ \Phi_{A} $ and the exchange probability of atom-$A$ in the bulk with atom-$B$ on the surface minus the inverse exchange probability of atom-$A$ on the surface with atom-$B$ in the bulk.   
On the other hand, due to the mass conservation for atoms and this fact that $ X^{b}_{A}(t)+ X^{b}_{B}(t)= 1 $ at any time the following conditions must be achieved \cite{Dehaes,Magri}: 
\begin{equation}
X^{s}_{A}(t)+ X^{b}_{A}(t)= X^{s}_{A}(0)+ X^{b}_{A}(0)+\Phi_{A}t
\end{equation} 
\begin{equation}
X^{s}_{A}(t)+ X^{s}_{B}(t)= X^{s}_{A}(0)+ X^{s}_{B}(0)+(\Phi_{A}+\Phi_{B})t
\end{equation}

\begin{figure} [htb]
\begin{center}
 \includegraphics[scale=0.43]{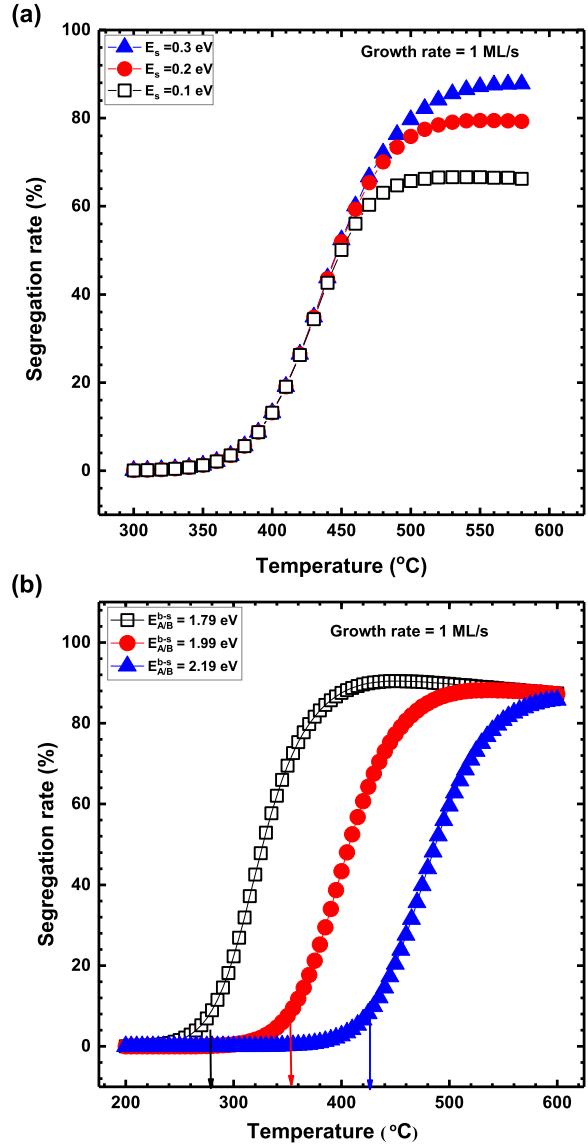}
 \end{center} 

\caption {(a)Variation of the segregation rate with respect to the growth temperature for a constant $E^{b\rightarrow s}_{In/Ga}$ of 1.8~eV and different segregation energies in the range of 0.1-0.3~eV. (b) Variation of the segregation rate with respect to the growth temperature for $E_{s}$= 0.2~eV and different  $ E^{b\rightarrow s}_{A/B} $ in the range of 1.79-2.19~eV. Atom-A segregation occurs at lower growth temperature with decreasing $ E^{b\rightarrow s}_{A/B} $. The arrows indicate the threshold growth temperature. }
\vspace{0.5 mm}
\label{fig:mesh.1}
\end{figure}

Using equations (4)-(6), it is possible to predict the concentration of atoms at any time for different growth conditions. The exchange rates, however, exponentially depend on the barrier energies, therefore, variations of the barrier energies alter the exchange rates drastically. This, in turn, changes the predicted surface concentration of atoms using the Kinetic Model. We investigated the impact of variation of $E_{s}$ and $E^{b\rightarrow s}_{A/B} $ on the segregation rate, which is defined as the ratio of the number of atoms-$A$ segregated to the surface to the total number of atoms-$A$ in the sub-surface layer. The results of our calculation indicate that for a constant $E^{b\rightarrow s}_{A/B} $= 1.8~eV \cite{Dehaes}, at high growth temperature, the segregation rate increases as $ E_{s} $ increases while at low growth temperature, the segregation rate remains unchanged with increasing the $ E_{s} $. Fig.\ref{fig:mesh.1}a shows that for the identical growth rate of 1~ML/s, the maximum segregation rate at higher temperature than 460~$^{\circ}$C is 68$\%$ for $E_{s}$ = 0.1~eV while, it increases to 80$\%$ for $E_{s} $= 0.2~eV and 88$\%$ for $E_{s} $= 0.3~eV. 

Variations of $ E^{b\rightarrow s}_{A/B} $ also affect the segregation rate. Variations of $ E^{b\rightarrow s}_{A/B} $ and $ E^{s\rightarrow b}_{A/B} $ for the constant $E_{s} $ = 0.2~eV result in the change of the growth temperature threshold for the segregation of atom-A. Fig\ref{fig:mesh.1}b depicts the results of segregation rate with respect to the growth temperature for an identical growth rate of 1~ML/s  and different $ E^{b\rightarrow s}_{A/B} $ and $ E^{s\rightarrow b}_{A/B} $. 
According to these results, the increase of $ E^{b\rightarrow s}_{A/B} $ from 1.79~eV to 2.19~eV leads to the increase of growth temperature threshold for atom- A segregation from 280~$ ^{\circ} $C to 427~$ ^{\circ} $C. 
Therefore, knowing the exchange ($ E^{b\rightarrow s}_{A/B} $) and segregation ($E_{s}$) energies, the concentration profile of the constituent atoms can be determined. 

\begin{figure} [htb]
\begin{center}
\includegraphics[scale=0.45]{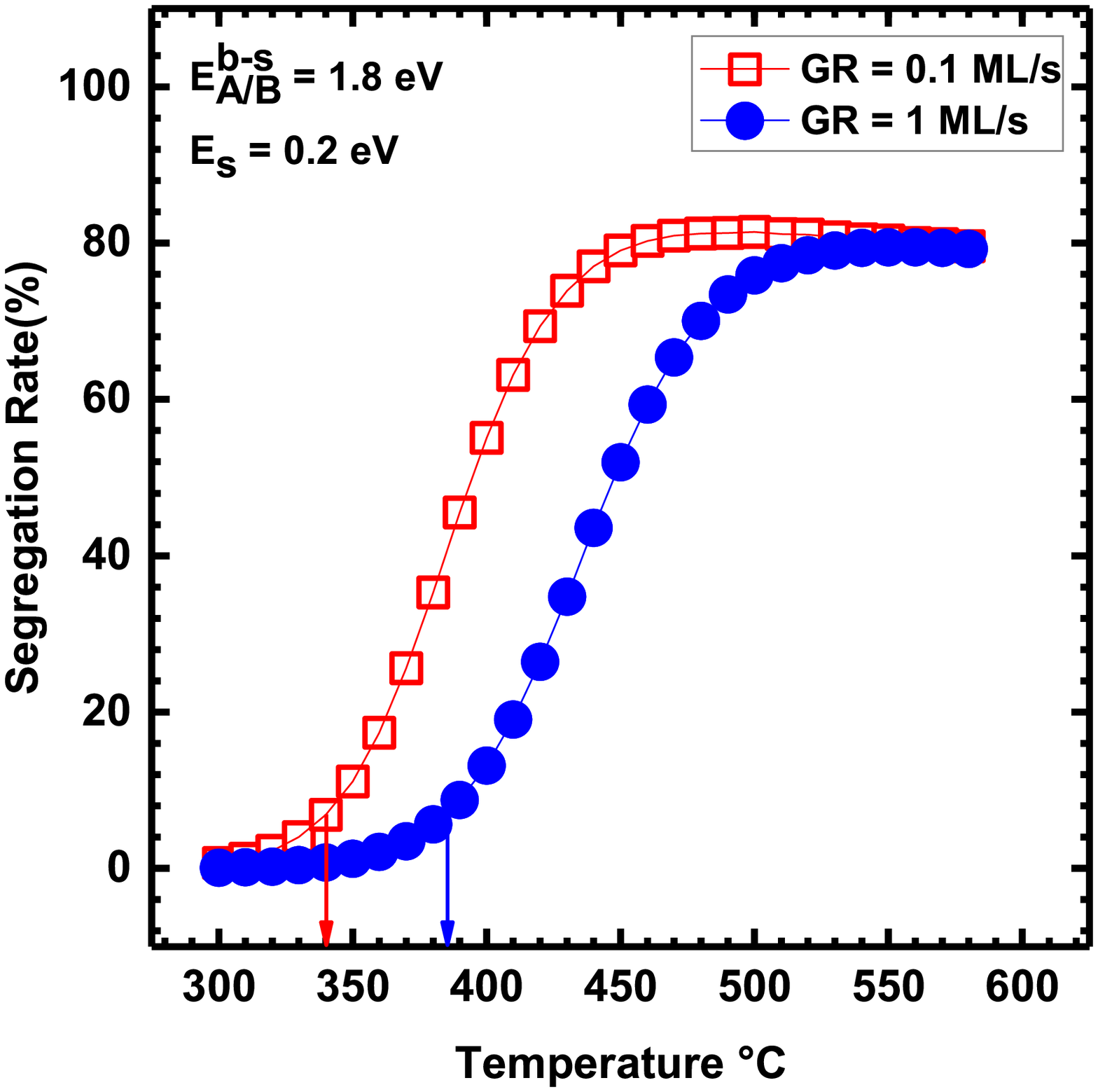} 
\end{center}
\caption {Variation of the segregation rate with respect to the growth temperature for $ E^{b\rightarrow s}_{A/B} $ =1.8~eV and $E_{s}$ = 0.2~eV for two different growth rates. The arrows indicate the growth temperature threshold for 5$ \% $ segregation rate. Reduction of the growth temperature threshold is seen with decreasing the growth rate. }
\label{fig:mesh.2}
\vspace{1 mm}
\end{figure}

Growth rate is another growth parameter, besides the growth temperature, that changes the segregation rate. For the constant $ E^{b\rightarrow s}_{A/B} $ = 1.8~eV and  $E_{s} $ = 0.2~eV, decrease of the growth rate from 1~ML/s to 0.1~ML/s is followed by the reduction of the threshold of growth temperature for the segregation from 380~$ ^{\circ} $C to 338~$ ^{\circ} $C (Fig\ref{fig:mesh.2}). According to these results, it is expected that the segregation of group-III elements occurs at the lower growth temperature for a lower growth rate and,therefore, to achieve minimum segregation the growth rate must be increased. 

\begin{figure} [h]
\begin{center}
\includegraphics[scale=0.35]{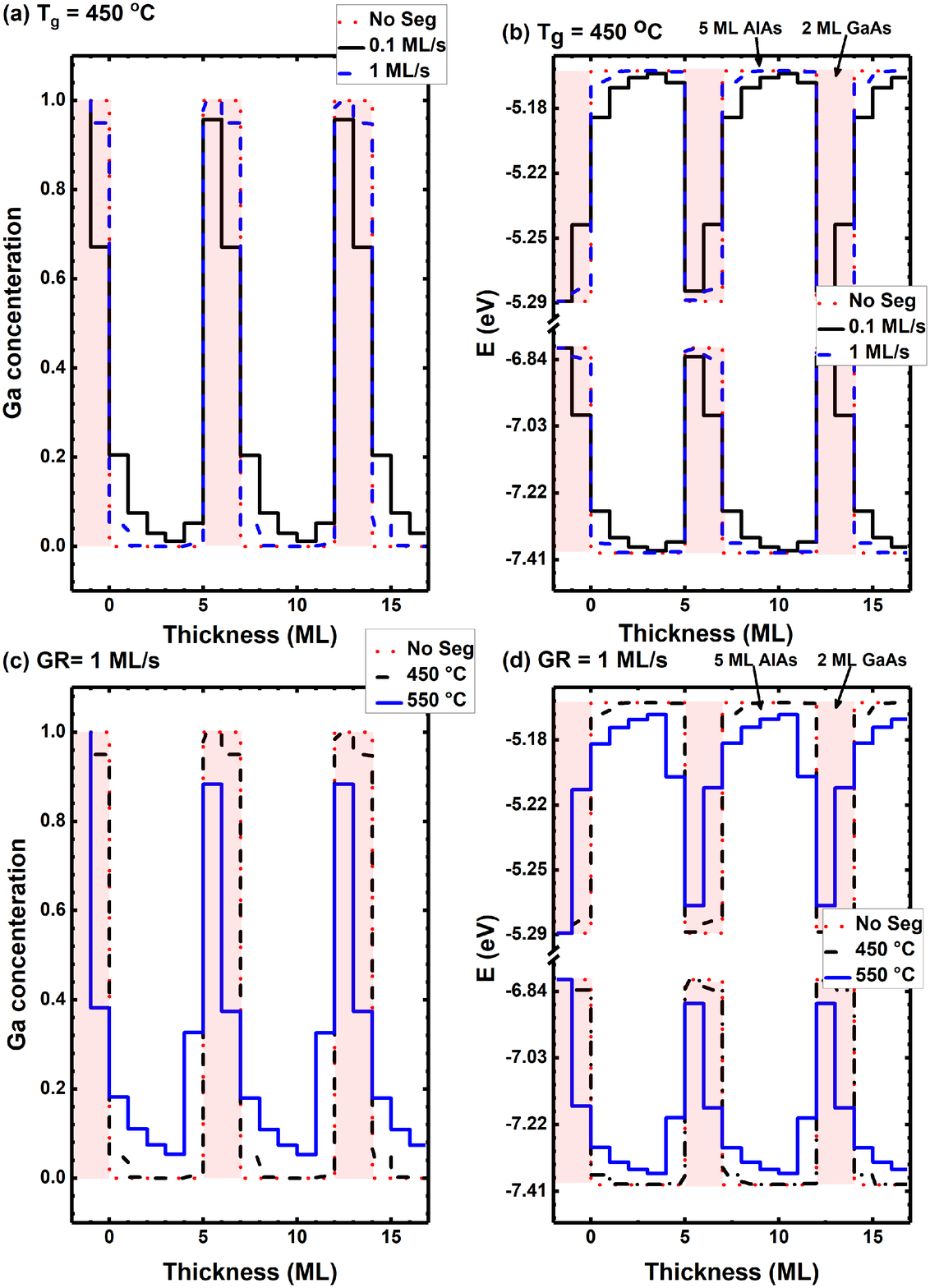} 
\end{center}
\caption {Ga concentration profile, (a) two growth rates (GR) of 0.1 and 1~ML/s at 450~$ ^{\circ}C $, and (c) GR = 1~ML/s and different growth temperatures (T$ _{g} $). Energy band lineup for 2~ML GaAs QW sandwiched between 5~ML AlAs barriers, (b) for T$ _{g} $ = 450~$ ^{\circ} $C and GR in the range of 0.1-1~ML/s,  (d) GR = 1~ML/s and different T$ _{g} $ of 450~$ ^{\circ} $C and 550~$ ^{\circ} $C. GaSb QW for the ideal case of non-segregation (No Seg) has been highlighted. }
\label{fig:mesh.3}
\end{figure}

Knowing that $ E^{b\rightarrow s}_{Ga/Al} $ = 2.0~eV \cite{Dehaes} and $ E_{s}  $ is 0.15~eV \cite{Moison,Kohleick} for AlGaAs system, we studied the impact of growth temperature and growth rate on the segregation of Ga atoms in a multi-QW system of 2~ML GaAs/5~ML AlAs on GaAs (001) surface. Fig.\ref{fig:mesh.3}a depicts the results of the Ga concentration versus thickness for two different growth rates (0.1 and 1~ML/s) at the growth temperature of 450~$ ^{\circ}C $. Interestingly, increase of the segregation due to the decrease of the growth rate from 1~ML/s to 0.1~ML/s changes the content of Ga atoms from 0.94 to 0.67 in the second ML of GaAs QW.
In the case of 1~ML/s growth rate, Ga atoms slightly segregate to the next AlAs layer and the segregation of the Ga atoms to the previous AlAs layer is not observed. However, larger segregation of Ga atoms at both next and previous AlAs layer is observed for the lower growth rate of 0.1~ML/s. The gradual variation of Ga (Al) content changes the energy band alignment of the quanum structure. Fig.\ref{fig:mesh.3}b illustrates the calculated energy band lineup for the non-segregated (highlighted) and segregated GaAs/AlAs system\cite{Walle}. The QW shape varies from a symmetrical to an asymmetrical shape without and with considering the segregation, respectively. The increase of the segregation of atoms with decreasing the growth rate form 1~ML/s to 0.1~ML/s modifies the GaAs QW width and AlAs barrier height. These modifications alter the optoelectronic properties of the desired quantum structure.

Similar behavior is obserevd for the Ga atoms profile at a given growth rate (1~ML/s) and different growth temperatures. The results for the growth at 450~$ ^{\circ} $C demonestrate that Ga atoms segregate from GaAs QW only to the next AlAs layer while the increase of the growth temperature up to 550~$ ^{\circ} $C results to a large intermixing of the Ga and Al atoms at the next and previous AlAs layer(Fig.\ref{fig:mesh.3}c). The content of Ga atom at the next and previous AlAs layer is 18$\%$ and 32$\%$, respectivley, when the growth temperature increses to 550~$ ^{\circ} $C, that proposes a larger intemixing of Ga and Al at the inverse interface of AlAs/GaAs. Fig.\ref{fig:mesh.3}d illustrates the energy band lineup for the non-segregated (highlighted) and segregated systems\cite{Walle}. A more significant change of energy band lineup is observed for a higher growth temperature of 550~$ ^{\circ}$C due to the larger migration of Ga atoms into the previous AlAs. Our results, therefore, demonstrate a stronger impact of the increase of the growth temperature on the atomic profile of the GaAs/AlAs quantum structure system in comparison with the reduction of the growth rate. Consequently, to grow a sample with the minimum migration of Ga atoms to the previous AlAs layer, the growth temperature must be lowered down to 450~$ ^{\circ} $C for the first ML growth of GaAs and afterwards it can be increased to the higher temperatures to improve the crystal quality and reduce the defects.

\begin{figure} [htb]
\begin{center}
\includegraphics[scale=0.45]{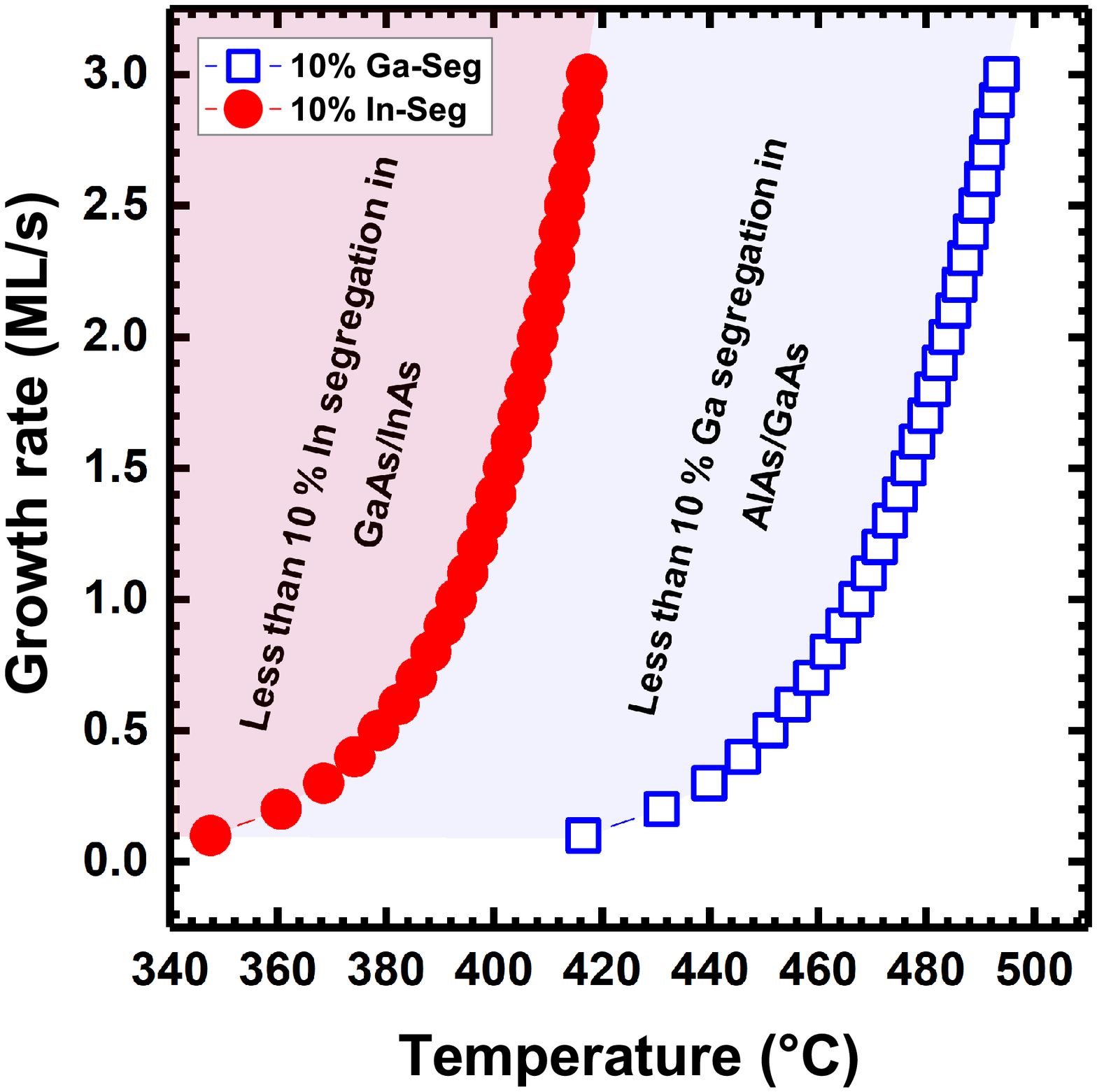}
\end{center}
\caption{Critical growth rate and growth temperature regions for the In and Ga  segregation rates $ \leq $ 10$\%$ in GaAs/InAs (red region) and GaAs/AlAs (red + blue regions), respectively. To grow heterostructures with less than 10$\%$ segregation rate, the growth rate must be increased with increasing the growth temperature.}
\label{fig:mesh.4}
\end{figure}

Considering the fact that the atoms segregation rate is a function of the growth temperature and the growth rate, the growth window, in which, the segregation rate is less than a certain percentage can be determined. For instance, Fig.\ref{fig:mesh.4} illustrates the growth rate and the growth temperature regions for GaAs/InAs and GaAs/AlAs heterostructures, in which the segregation rate is less than 10$\%$. Likewise, we can calculate the growth temperature and growth rate for any given segregation rate. The Kinetic Model results imply that the growth temperature threshold for the segregation of Ga atoms at the normal interface into the next AlAs layer occurs at a higher temperature in comparison with the In atoms segregation in to the adjacent GaAs layer. This is related to the lower barrier energy for In segregation from bulk to the surface ($ E^{b\rightarrow s}_{In/Ga} $ = 1.8~eV) in GaAs/InAs heterosystem compared with the barrier energy for Ga segregation ($ E^{b\rightarrow s}_{Ga/Al} $ = 2.0~eV) in AlAs/GaAs system.   Thus, to grow the GaAs/InAs and GaAs/AlAs heterostructures with segregation rate of less than 10$\%$ and same growth rate of 1~ML/s, the growth temperature must be lowered form 468$ ^{\circ} $C for GaAs/AlAs to 394$ ^{\circ} $C for GaAs/InAs system. 

In conclusion, our results show the critical boundary of the barrier energy of $ E^{b\rightarrow s}_{A/B}$ on the threshold growth temperature and growth rate for the segregation of atom-A from the bulk to the surface. We have calculated the segregation rate of Ga atoms for the GaAs/AlAs system. Our results indicate the increase of the segregation rate with increasing the growth temperature and decreasing the growth rate. The energy band lineup calculations, on the other hand, show the change of the QW from a symmetric to an asymmetric QW due to the gradual variation of the Ga content in the GaAs/AlAs quantum structure. These results demonstrate the stronger impact of the growth temperature compared with the growth rate on the Ga segregation profile. We also determined the growth window for the growth temperature and the growth rate to achieve less than 10$\%$ segregation rate of In and Ga atoms in the (In,Ga)As and (Al,Ga)As material systems, which can be very useful from the practical point of view. 


\end{document}